%% file: main_body.tex
\documentclass[fleqn,usenatbib]{mnras}

\usepackage{amsmath}

\usepackage{newtxtext,newtxmath}
\usepackage[T1]{fontenc}

\DeclareRobustCommand{\VAN}[3]{#2}
\let\VANthebibliography\thebibliography
\def\thebibliography{\DeclareRobustCommand{\VAN}[3]{##3}\VANthebibliography}

\usepackage{graphicx}	% Including figure files
\usepackage{amsmath}	% Advanced maths commands
\title[The satellite galaxies around M31]{The lives and deaths of faint satellite galaxies around M31}

\author[A. Merrow et al.]{
Alex Merrow,$^{1,2}$\thanks{E-mail: a.j.cooke@2022.ljmu.ac.uk}
Kyle A. Oman,$^{3,4}$
Azadeh Fattahi$^{3,5}$
\\
$^{1}$Astrophysics Research Institute, Liverpool John Moores University, 146 Brownlow Hill, Liverpool, L3 5RF, UK\\
$^{2}$Department of Physics, Durham University, South Road, Durham DH1 3LE, UK\\
$^{3}$Institute for Computational Cosmology, Department of Physics, Durham University, South Road, Durham DH1 3LE, UK\\
$^{4}$Centre for Extragalactic Astronomy, Department of Physics, Durham University, South Road, Durham DH1 3LE, UK\\
$^{5}$The Oskar Klein Centre, Department of Physics, Stockholm University, Albanova University Centre, 106 91 Stockholm, Sweden\\
}

\date{Accepted XXX. Received YYY; in original form ZZZ}

\pubyear{2025}

\begin{document}
\label{firstpage}
\pagerange{\pageref{firstpage}--\pageref{lastpage}}
\maketitle

\begin{abstract}
We present predictions for proper motions, infall times and times of first pericentric passage for 39 of M31's satellite galaxies. We estimate these by sampling satellite orbits from cosmological N-body simulations matched on mass, distance and velocity along the line of sight, in addition to properties of the host system. Our predictions are probabilistic based on repeated sampling from the uncertainty distributions of all quantities involved. We use these constraints on the satellites' orbital histories in conjunction with their published star formation histories to investigate the dominant environmental mechanisms for quenching satellites of M31-like hosts. Around half of the satellites appear to have quenched before their first pericentric passage around M31. Only the most massive satellites (with stellar masses $>10^8\mathrm{M}_{\odot}$) are able to maintain star formation for up to billions of years after infall. The majority of faint satellites, with $M_\star < 10^8\,\mathrm{M}_\odot$, were likely quenched before entering the M31 system. We compare our results for M31 against predictions for the Milky~Way's satellites from the literature; M31's has a more active recent accretion history with more recently quenched satellites than the Milky~Way.
\end{abstract}


%%%%%%%%%%%%%%%%%%%%%%%%%%%%%%%%%%%%%%%%%%%%%%%%%%

%%%%%%%%%%%%%%%%% BODY OF PAPER %%%%%%%%%%%%%%%%%%

\section{Introduction}
\label{intro}

The Local Group offers a unique testbed for galaxy evolution theories due to our unique vantage point within it. For the Milky~Way in particular, the advent of high-precision proper motion measurements across much of the Galactic system has enabled reasonably accurate integration of the orbits of satellites back in time \citep{2018Si,2018Fr,2018Fr2,2024Sa}. Alongside this, the ability to spatially resolve individual stars allows for the determination of precise star formation histories for the same objects \citep{1994SH,2002Do,2008DJ,2014We,2021Sa}. Combining both of these measurements with the paths of satellites in cosmological simulations can lead to insights into the physical processes shaping the evolution of satellites as they orbit their hosts \citep{2012Ro,2019Fi}.

\citet{2012Ro} used the subhaloes of a single simulated Milky~Way-mass dark matter halo to find infall times for 21 of the Milky~Way's satellites by comparing radial distance and velocity (proper motions were available and utilised for 7 of these) between observed satellites and the simulated subhaloes at $z=0$. For those satellites with well constrained infall times, they found that the classical satellites are typically quenched after infall, while the ultra-faint dwarfs had quenched significantly before infall onto the Milky~Way. More recently, \citet{2019Fi} used 12 simulated Milky~Way-mass haloes to match subhaloes' normalised binding energies, normalised distances from their hosts, and direction of radial velocities to those of the Milky~Way's satellites. With detailed star formation histories then available for 15 of these satellites, they found that satellites with stellar mass $10^5 < M_{\star}/\mathrm{M}_\odot < 10^8$ are consistent with having quenched only shortly after infall, while satellites with masses lower than this quenched earlier.

These results are broadly in line with the mechanisms that we expect to drive the quenching of dwarf galaxies now existing in dense environments. Ram-pressure stripping acts as a galaxy moves quickly through a dense medium, removing the gas required to form new stars from the outside in \citep{1972Gu,1976Gi,1976Le}. This can occur at any point during a dwarf galaxy's evolution before becoming a satellite of a larger galaxy, for example when encountering dense filaments in the cosmic web, but is particularly prominent while infalling through the circumgalactic medium surrounding a new host \citep{1999Mu,2003Ma,2006He,2013Be}. Tidal stripping is instead more prominent at pericentre passage shortly afterwards, where the tidal forces from a host galaxy are strongest. This strips the outer layers first, including the dark matter halo, but also the star-forming gas in many cases \citep{1975Ri,1983Me,2006Re,2016Sm}. Both of these mechanisms are most effective with lower-mass satellite galaxies, since a shallower potential well binds gas more weakly. The accretion of new gas is also slowed/halted by the host environment \citep{1980La,2000Ba,2015Pe,2025ViZa}. These mechanisms are consistent with the evolution of the intermediate stellar mass Milky~Way satellites studied by \citet{2019Fi} which quench shortly after infall, with pre-satellite ram-pressure stripping being at least partially responsible for the lower mass, early quenching satellites present in both mentioned studies.

Regardless of environment or orbital history, reionization is thought to be responsible for quenching many low-mass galaxies around $10\,\mathrm{Gyr}$ ago \citep{2000Bu,2005Ri,2014Br}, contributing to the population of low-mass satellites that quenched well before infall. Internal mechanisms such as supernova (SN) or active galactic nuclear (AGN) feedback can suppress star formation across all galaxy masses, and quench even massive dwarf galaxies over a long timescale \citep{1974La,1986De,2003Be,2005DM,2006Cr,2012Fa,2014Ho}, giving a potential mechanism for the post-infall quenching of the classical Milky~Way satellites as proposed by \citet{2012Ro}.

The detailed studies enabled by the wealth of information available for the Milky~Way system trades off against the fact that it is a single host galaxy, raising questions about the generality of any conclusions drawn. Studies of external host galaxies and their satellites are limited by the available information, usually position offsets projected onto the plane of the sky, velocity offsets projected along the line of sight, and star formation histories based on integrated stellar light \citep[e.g.][]{1993Za,2003Pr,2011Ka,2012Wa}. This approach has clear advantages in terms of sample size and diversity of host galaxy properties, although samples are inevitably limited to brighter satellites than the faintest ones observed around our own Galaxy.

The other massive galaxy in the Local Group, M31, provides an interesting intermediate case. Reliable distances, line-of-sight velocities, and star formation histories based on resolved stellar populations are available for almost all of M31 satellites \citep{2012MC,2022Sa,2022Na,2009Wi,2014We,2017Sk,2019We,2023Sa,2025Sa}. Moreover, proper motions are available for a subset of them \citep{2005Br,2007Br,2020So,2024Ru,2025CD,2025Be}, albeit with lower precision than can be achieved in the Milky~Way system. The M31 satellite system is particularly attractive to compare to that of the Milky~Way since, while the two host galaxies are broadly similar (e.g. mass, morphology, etc.), there is mounting evidence that their accretion histories are very different with M31 being much more active recently in this regard \citep[e.g.][]{2004Ib,2006Fa,2006Fo,2018Ha,2018MC,2018DS,2023De} -- a finding that we will corroborate in this paper.

In this work, we use a compilation of observed phase space coordinates and photometry of resolved stars for the satellites of M31 (Sec.~\ref{obs}) and a statistical sample of satellite orbits drawn from an N-body simulation (Sec.~\ref{sim}). Matching satellites to simulated satellite haloes by their 3D positions and line-of-sight velocities allows us to predict their proper motions and parameters describing their likely orbits. We derive relationships between their orbital and star formation histories (Sec.~\ref{mainresults}) and discuss these in the context of a comparison with the Milky~Way satellite system (Sec.~\ref{mw-comparison}) and findings from other works (Sec.~\ref{literature}).

\section{Methodology}

\subsection{Observations}
\label{obs}

\citet{2012MC} provides a compilation of observational data for dwarf galaxies in the Local Group, including satellites of M31, which has been kept up to date until 2021 with recent measurements. From this compiled table, we take each galaxy's right ascension, declination, heliocentric radial velocity ($v_{\mathrm{h,sat}}$), distance modulus ($\mu$), and apparent magnitude ($m$). We supplement these with more recent distance estimates from \citet{2022Sa}. We select satellites of M31 by making a cut in distance from M31 of 2 virial radii and a cut in velocity relative to M31 of 1.5 times the 3D velocity dispersion of satellites, giving us 39 satellites in total (see below for the values taken for M31's virial radius and 3D velocity dispersion). Since only a quarter of these currently have proper motions available (Andromeda~III, Andromeda~V, Andromeda~VI, IC~10, NGC~147, NGC~185, LGS~3, PegasusdIrr and Triangulum), we base our analysis on the line of sight velocity measurements for consistency across the satellites.

For quenching times, we use the star formation rates of 37 of M31's satellites\footnote{Andromeda~XVIII and Andromeda~XXVII have kinematic information, but no star formation history available.} from \citet{2014We,2019We,2025Sa}. The latter two papers directly provide the lookback times with uncertainties to $\tau_{90}$ (the lookback time at which the galaxy has formed 90~per~cent of its total stellar mass) which we will use as a proxy for quenching time in this paper. In \citet{2014We}, we instead calculate these values from the cumulative star formation histories provided. It should be noted that even for a continuously star forming galaxy, $\tau_{90}$ will give a `quenching time' in the past. From morphological classifications in \citet{2012MC}, 5 of the 37 M31 satellites are unlikely to be quenched (these are IC~10, Triangulum, LGS~3, IC~1613, and PegasusdIrr).

Lastly, for M31 itself, we use observed values of distance $d_\mathrm{M31}=752\pm27\,\mathrm{kpc}$ from \citet{2012Re}, heliocentric radial velocity of $v=-300\pm4\,\mathrm{km}\,\mathrm{s}^{-1}$ from \citet{2016Sa}, and virial halo mass $M_\mathrm{vir}=(1.54\pm0.39)\times10^{12}\mathrm{M}_{\odot}$ from \citet{2012VM}. This virial mass definition matches that which we adopt in our analysis of our simulation (Sec.~\ref{sim}): the mass inside of the sphere within which the mean density is 360 times the background matter density \citep{1998Br}. This corresponds to a virial radius of $308\pm26\,\mathrm{kpc}$. From the virial mass, we also estimate the 3D velocity dispersion ($\sigma_\mathrm{3D}$) of dark matter particles in the galaxy as:
\begin{equation}
    \sigma_\mathrm{3D} = 0.0165\left(\frac{M_\mathrm{vir}}{\mathrm{M}_{\odot}}\right)^{\frac{1}{3}}
\end{equation}
derived from equation (2) in \citet[][see also \citealp{2013Mu}]{2006Bi}, giving $\sigma_\mathrm{3D}=180\pm15\,\mathrm{km}\,\mathrm{s}^{-1}$.

\subsubsection{Constructing a distribution of observational parameters}
\label{obsprocessing}

In order to match observed satellite galaxies to relevant simulated satellite haloes below, we express the observational data and their uncertainties in the form of probability distributions. We model each observed value as a joined pair of half-normal distributions. These consist of the left half of a normal distribution centred at the observed value, with width given by the lower uncertainty, combined with the right half of a normal distribution centred at the same observed value, but with width given by the upper uncertainty. We normalise each half to have an area of 0.5. From this, we obtain a (discontinuous) full probability distribution whose $50^{\mathrm{th}}$, $16^{\mathrm{th}}$ and $84^{\mathrm{th}}$ percentiles match the observation and its lower and upper uncertainties, respectively.

From each observation's distribution (across all galaxies), we randomly and independently select a single value. Using these we calculate three values for each satellite: a 3D distance to M31, calculated from $\mu$, $d_{\mathrm{M31}}$ and position on the sky; a signed line of sight velocity offset from M31, calculated from $v_{\mathrm{h,sat}}$ and $v_{\mathrm{h,M31}}$; a stellar mass, calculated from $\mu$ and $m$ while assuming a stellar mass-to-light ratio of $1.5\,\mathrm{M}_{\odot}\,\mathrm{L}_{\odot}^{-1}$.

We also select a random value of $M_\mathrm{vir}$ (and derive the corresponding values for $r_\mathrm{vir}$ and $\sigma_\mathrm{3D}$) from the distribution for M31. With these, we normalise radial separation from M31 by $r_{\mathrm{vir}}$ and normalise the velocity offset by $\sigma_{\mathrm{3D}}$. Our normalised distance from M31 is then given by $r_{\mathrm{3D}}/r_{\mathrm{vir}}$ where $r_{\mathrm{3D}}$ is the satellite's distance from M31, while our normalised signed velocity relative to M31 is given by:
\begin{equation}
    \frac{V_{\mathrm{LoS}}}{\sigma_{\mathrm{3D}}} =
    \begin{cases}
    \frac{v_{\mathrm{h,sat}}-v_{\mathrm{h,M31}}}{\sigma_{\mathrm{3D}}} & \mathrm{if} d_{\mathrm{h,sat}} > d_{\mathrm{M31}} \\
    \frac{v_{\mathrm{h,M31}}-v_{\mathrm{h,sat}}}{\sigma_{\mathrm{3D}}} & \mathrm{if} d_{\mathrm{h,sat}} < d_{\mathrm{M31}}
    \end{cases}
\end{equation}
where $d_{\mathrm{h,sat}}$ is the satellite's heliocentric distance. These two cases fix the velocity's sign to be positive if the satellite is moving away from M31 along the line of sight.

In addition to stellar mass, we also estimate the peak virial halo mass for each satellite galaxy. \citet{2019Be} gives a model for the correlation in galaxies between stellar mass and the peak halo mass, $M_\mathrm{peak}$. It is however worth noting, due to incomplete samples (both observational and simulated) at lower stellar masses, that the model must be extrapolated for dwarf galaxies where $M_\mathrm{peak}<10^{11}\mathrm{M}_{\odot}$. Thus the halo mass estimates for our observed satellites carry large uncertainties. In addition, the satellites that we study may have been substantially stripped of stars by M31. The model does not account for this, resulting in a potential underestimate of their peak halo mass. We numerically invert the equation:

\begin{equation}
    \label{Behroozi}
    \log_{10}\left(\frac{M_{\star}}{M_{1}}\right) = \epsilon - \log_{10}\left(10^{-\alpha x}+10^{-\beta x}\right) + \gamma \exp\left(-0.5\left(\frac{x}{\delta}\right)^2\right),
\end{equation}

 where $x = \log_{10}\left(M_\mathrm{peak}/M_{1}\right)$, to obtain an estimated peak halo mass from the stellar mass of M31's satellites. The parameters in this equation depend on redshift, but we omit this dependence and set $z=0$ due to M31's close proximity, giving a set of 6 constants with values and uncertainties given by \citet{2019Be}. We treat each of these constants as an additional variable and model a distribution for each to select a value from, as described above for observations.

The above describes the process for a single value chosen for each relevant variable. We repeat this for 10,000 iterations to obtain a final 10,000 normalised values of $r_{\mathrm{3D}}/r_{\mathrm{vir}}$, $V_{\mathrm{LoS}}/\sigma_{\mathrm{3D}}$ and $M_{\mathrm{peak}}$ for each satellite, and treat this as a discrete approximation to the covariant probability distributions of these quantities.

\subsection{Simulation}
\label{sim}

To compare with our compiled observations, we require a large catalogue of satellite galaxies with cosmologically-motivated orbital histories and some information on future evolution. To this end we have extended the dark matter-only version of the EAGLE simulation \citep[][run `L100N1504' in their notation]{2015Sc} to a final scale factor of 2 (about $10\,\mathrm{Gyr}$ into the future). The simulation consists of $1504^3$ dark matter particles of mass $9.70\times10^6\,\mathrm{M}_\odot$ evolving in a $100\,\mathrm{cMpc}$ periodic box with softening length $2.66\,\mathrm{ckpc}$ (up to a maximum of $0.70\,\mathrm{kpc}$ from $z=2.8$ onwards). The cosmology used, observationally motivated by \citet{2014Pl}, takes $h=0.677$, $\Omega_{\mathrm{m}}=0.355$, and $\Omega_{\mathrm{\Lambda}}=0.693$. We identify haloes using the \textsc{rockstar} halo finder \citep{2012Be} and create merger trees using the \textsc{consistent-trees} tool described in \citet{2012Be2}.

With these tools we obtain the 3D position and velocity in the comoving simulation box and virial mass for each simulated satellite and each simulated host, all recorded at each time step for which the satellite exists. With this combination of measurements we have enough information to measure projected phase space coordinates for the simulated satellites ($r_{\mathrm{3D}}/r_{\mathrm{vir}}$, $V_{\mathrm{LoS}}\sigma_{\mathrm{3D}}$), and $M_{\mathrm{peak}}$, matching the parameters obtained for observed satellites in Sec.~\ref{obs}. We take the line of sight as the $z$-axis of the simulation box. We also record the time at which each satellite first crossed within $1\,r_\mathrm{vir}$ of the centre of the host halo (hereafter referred to as the infall time, $t_\mathrm{infall}$) and the time of the satellite's first pericentre passage around its host following infall (hereafter referred to as the pericentre time, $t_\mathrm{peri}$). The choice of these two times is motivated by the main environmental quenching mechanisms expected around M31 as outlined in Sec.~\ref{intro}: tidal stripping is most prominent during a satellite's pericentre passage, with the first such event being at $t_\mathrm{peri}$, while ram-pressure stripping likely gradually starts around $t_\mathrm{infall}$. For infall times we linearly interpolated between the last snapshot outside $r_\mathrm{vir}$ and the first snapshot inside $r_\mathrm{vir}$, while for pericentre times we recorded the first snapshot\footnote{The snapshot output cadence is variable but is typically about $200$~Myr, and never more than $400$~Myr.} after pericentre, since satellite motion is non-linear at this time. Lastly, we record the remaining components (those perpendicular to the `line of sight' $z$-axis) of each satellite's velocity relative to its host galaxy, decomposed into projected radial velocity away from the host, $v_\rho$, and the magnitude of the projected velocity perpendicular to this, $v_\phi$.

We restrict our simulated sample to satellites with $10^8 < M_{\mathrm{peak}}/\mathrm{M}_{\odot} < 10^{12}$, $0 < r_{\mathrm{3D}}/r_{\mathrm{vir}} < 2.5$, $-2 < V_{\mathrm{LoS}}/\sigma_{\mathrm{3D}} < 2$, and $5\times10^{11} < M_{\mathrm{vir,host}}/\mathrm{M}_{\odot} < 5\times10^{12}$. These restrictions ensure similarity to the M31 system while including a wide enough region to accommodate statistically useful samples for all of M31's observed satellites. The mass restriction also excludes the lowest-mass satellites which dominate the simulated population but are not seen in current observations.

\begin{figure*}
	\includegraphics[width=\textwidth]{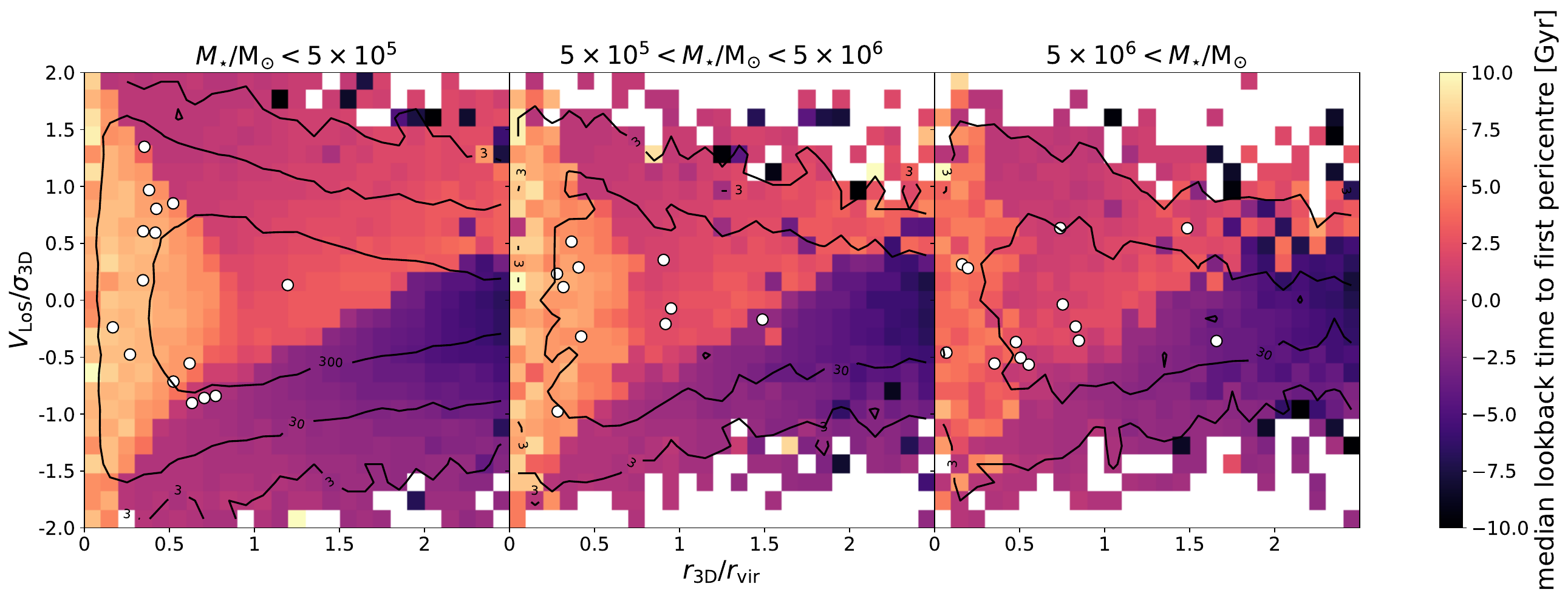}
    \caption{The dependence of first pericentre time on our chosen parameter space within the simulation. 3D radial distance from the host is on the horizontal axis (normalized by $r_\mathrm{vir}$), line of sight velocity relative to the host is placed on the vertical axis (normalized by $\sigma_\mathrm{3D}$), and satellite stellar mass is indicated by the 3 panels of low, middling, and high masses, as labelled. The space is coloured by the median lookback time to the first pericentre passage, ranging from ancient times in white/yellow to future times (negative values) in black/purple. We show contours of the number of simulated satellites per bin at 3, 30, and 300. The white points show the parameters (most probable values) for each observed satellite of M31.}
    \label{fig:allsim}
\end{figure*}

Fig.~\ref{fig:allsim} shows the distribution of our sample in this space projected onto the $r_{\mathrm{3D}}/r_{\mathrm{vir}}$--$V_{\mathrm{LoS}}/\sigma_{\mathrm{3D}}$ plane and separated into three stellar mass bins. From left to right these panels cut the space into a low mass bin, a middling mass bin and a high mass bin. Within each panel the contours show the distribution of simulated satellites in phase space and the white points indicate the corresponding positions of our sample of 39 satellites of M31. The density of the simulated space peaks at low velocity offset and at a radial distance of around $0.7\,r_\mathrm{vir}$. The higher count contour on the left panel indicates a higher number of satellites at low masses. Our sample of observed satellites is generally well spread across this distribution, giving a number of suitable matches: $100-6500$ for each satellite, except for Triangulum with just $10$  matches due to its high mass.

Fig.~\ref{fig:allsim} also shows the dependence of median first pericentre time on these parameters, by the colour scale. Each panel consists of three sharply split regions. At negative relative velocity (towards the host) and excluding low radii is the infalling population in black/purple. These satellites are approaching the host galaxy for the first time at the present day and as such have future (negative) pericentre times. The backsplash population in magenta/orange exists at mostly positive relative velocity and also typically outer radii. These are the satellites that have already experienced their first pericentre but are still on their first full orbit around the host galaxy at the present day. Lastly we have the ancient satellite population in orange/yellow at inner radii across the full range of velocities. These are the satellites that experienced their first pericentre passages long ago and have had one or more additional passages since, now existing on orbits which no longer take them far from the host galaxy. A typical satellite will move from the infalling region, through the backsplash region via flipping its signed velocity upon pericentre, finally settling in the ancient region close to its host. While the different groups mix significantly in this projected phase space, it is clear that radius and signed velocity give meaningful information on what point a satellite's orbit is at during the present day and thus how long ago it experienced major milestones in its orbit. The sign of the velocity is particularly valuable, c.f. \citet[][fig.~4]{2013Om}. The third dimension of our parameter space, peak halo mass, provides additional information on a satellite's progress through its orbit, with the ancient population decreasing in density at higher masses in exchange for a higher proportion of the backsplash population. This is likely because dynamical friction acts more strongly on satellites with masses closer to that of their host \citep{1943Ch,2008BK}, therefore much of the population of earlier infalling, more massive satellites - which would otherwise occupy this region of somewhat settled orbits - have instead merged with, or been disrupted by, their host before $z=0$.

\subsection{Orbital parameter inference}
\label{distributions}

\begin{figure*}
\centering
\begin{minipage}{.33\textwidth}
  \centering
  \includegraphics[width=\linewidth]{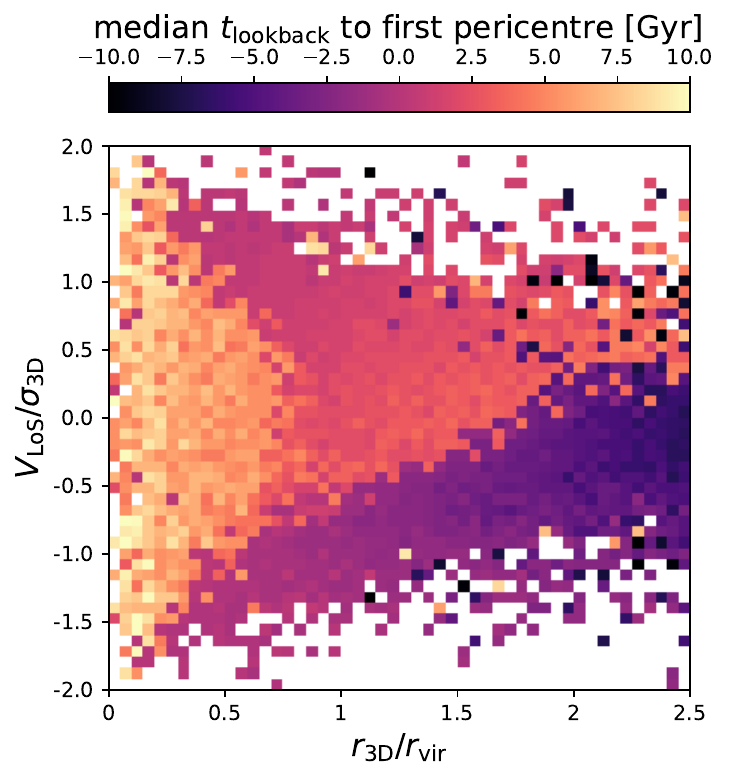}
\end{minipage}%
\begin{minipage}{.33\textwidth}
  \centering
  \includegraphics[width=\linewidth]{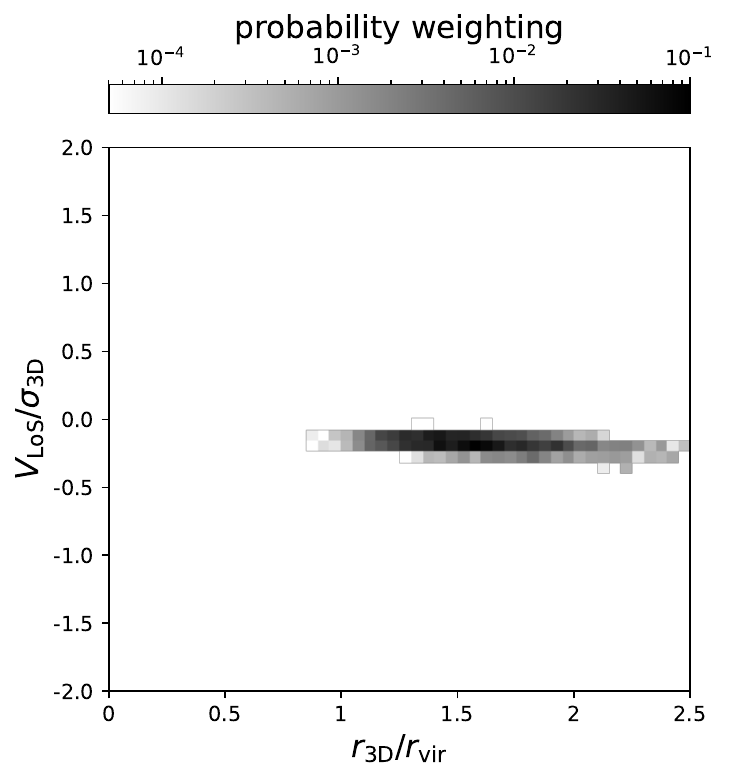}
\end{minipage}
\begin{minipage}{.33\textwidth}
  \centering
  \includegraphics[width=\linewidth]{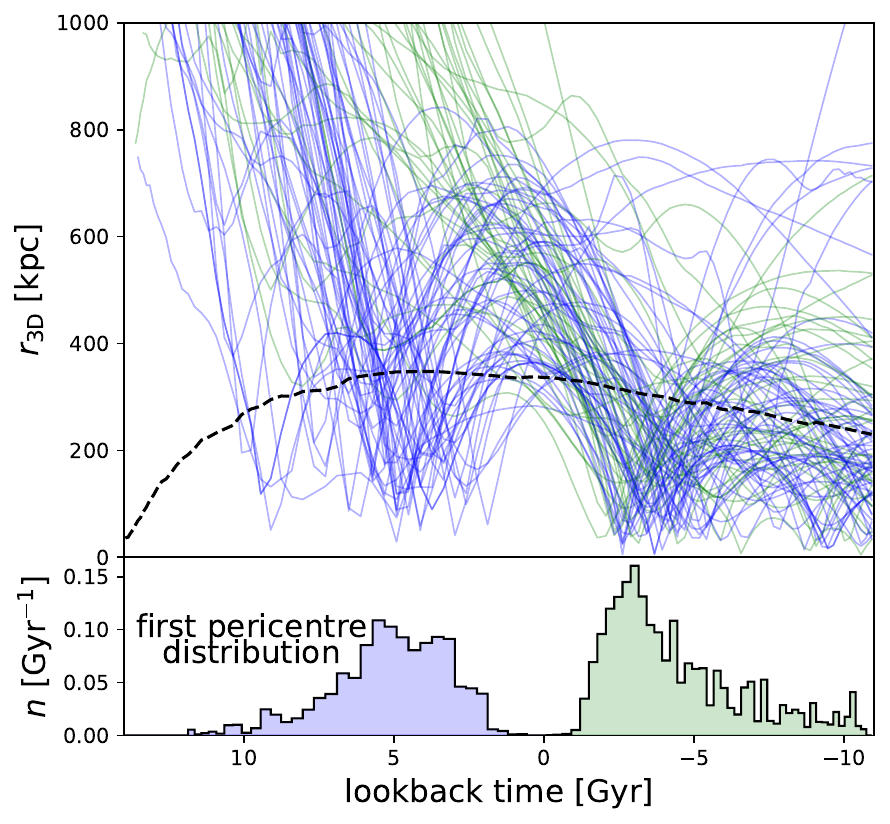}
\end{minipage}
\begin{minipage}{.33\textwidth}
  \centering
  \includegraphics[width=\linewidth]{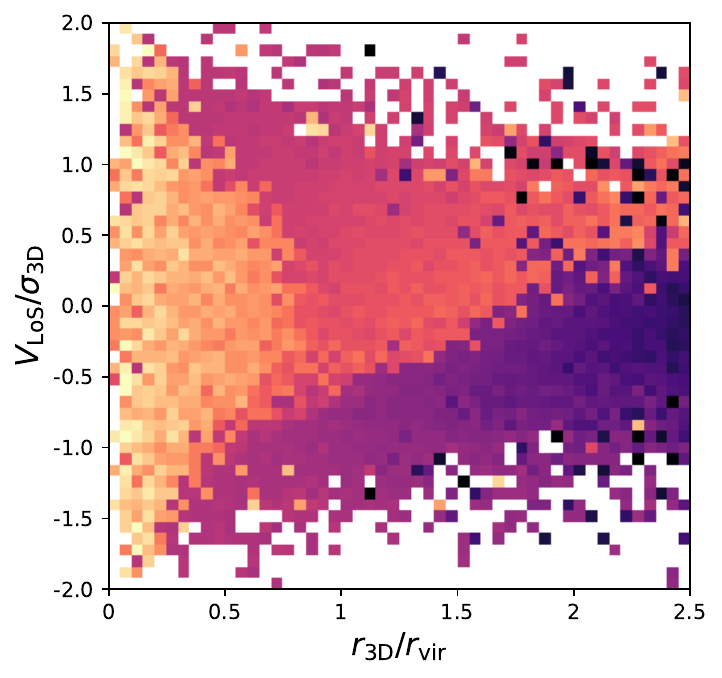}
\end{minipage}%
\begin{minipage}{.33\textwidth}
  \centering
  \includegraphics[width=\linewidth]{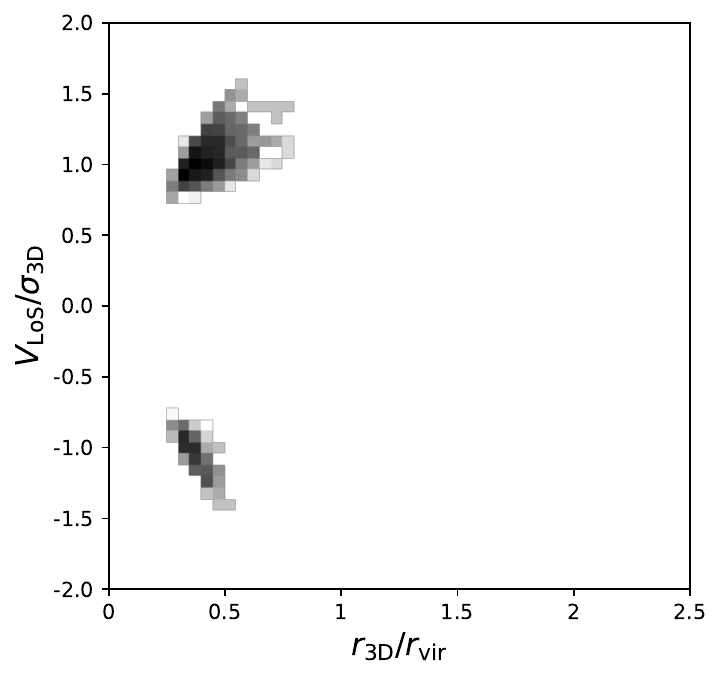}
\end{minipage}
\begin{minipage}{.33\textwidth}
  \centering
  \includegraphics[width=\linewidth]{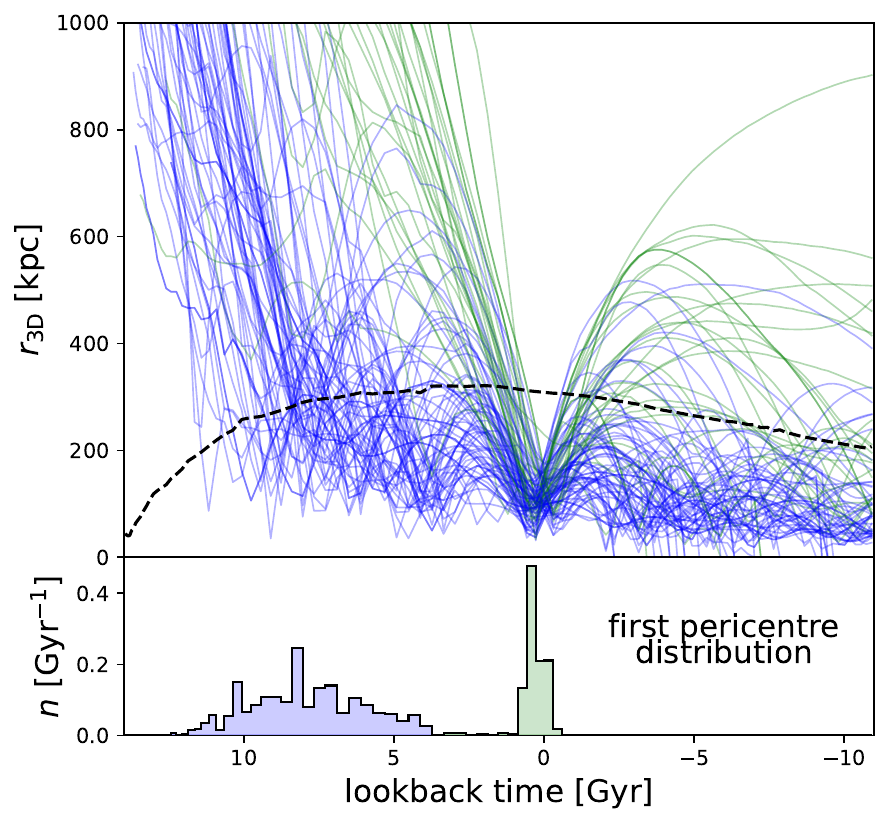}
\end{minipage}
\caption{Illustration of the method for predicting orbital parameters for examples Andromeda~XVIII (top) and Andromeda~XIX (bottom). The left panels show the dependence in the simulation of pericentre times on phase space as in Fig.~\ref{fig:allsim} but restricted to halo masses within a factor of 2 of the observed satellite. The middle panels show the probability distributions of the observed distance and line of sight velocity of the satellites, with a logarithmic colour scale. These provide a weighting to select the simulated haloes shown in the right panels. In these panels, we show comoving distance from the host for 100 randomly selected (with weighting) simulated satellites, with the lower histograms depicting the resultant spread of first pericentre times. The green and blue paths correspond to the later and earlier peaks in the histogram, respectively. For reference, the black dashed line in this panel shows the median virial radius of the hosts at each timestep.}
\label{fig:examples}
\end{figure*}

Having obtained values of $r_{\mathrm{3D}}/r_{\mathrm{vir}}$, $V_{\mathrm{LoS}}/\sigma_{\mathrm{3D}}$, and $M_{\mathrm{peak}}$ for both our observed and simulated satellite samples, we combine the two in order to relate them and their properties. We divide each dimension of the 3D parameter space into 50 bins, producing a grid with 125,000 cells to contain the 10,000 realisations for each satellite and 175,613 relevant simulated satellites. This bin spacing is chosen to be both fine enough to capture the detailed behaviour of both observed and simulated distributions, while being coarse enough that most bins in the denser regions are populated by both types of point to allow an association to be made. For each of M31's satellites we can then give a weighting to each simulated halo given by the number of points sampled from the observational uncertainty distributions contained in its bin, divided by the number of simulated haloes in the same bin. Once normalised to sum to 1 across all simulated satellites, this weighting represents the probability that the real satellite's properties best match those of the specific simulated halo.

As well as our 3 main parameters, each simulated satellite halo also carries an infall time, pericentre time, and proper motion as described in Sec.~\ref{sim}, for which we do not have corresponding values from observations. However, from the weightings that we obtain as described above, we create a probability distribution for each, constructed from the simulated haloes' properties.

Fig.~\ref{fig:examples} illustrates this process for the pericentre times of Andromeda~XVIII and Andromeda~XIX. The left panels are similar to Fig.~\ref{fig:allsim}, but limit the simulated haloes shown to those with peak mass within a factor of two of the observed satellite's peak halo mass. (This selection is for illustration purposes only and is not used in constructing any probability distributions.) Additionally, the bins shown are now the size of those used for our weighting. White, empty bins only exist at the extremities of the space, so that each bin in the region populated by M31's satellites contains simulated haloes to compare to.

The middle panels show the weighting resulting from the distribution of the satellites' 10,000 sampled points in the same parameter space. These show some typical features of these weightings. Due to a difference in the observational uncertainties our weighting for Andromeda~XVIII is more spread in the distance direction than the velocity direction, while for Andromeda~XIX the grey clusters are more even. However, Andromeda~XIX's distribution is split into two parts in phase space, one at positive and one at negative velocity. This appears in many of the satellites closest to M31 and is a result of small separation in the line of sight direction. In these cases, the uncertainty in distances along the line of sight is enough that some of the 10,000 samples place the satellite in front of M31 while some place it behind, thus flipping the sign of the velocity and populating two distinct regions in this projected phase space. In addition, there is a correlation between distance and velocity in both panels despite independently sampling each observational value. This is a result the correlation of the two normalisation factors: both $r_{\mathrm{vir}}$ and $\sigma_{\mathrm{3D}}$ increase with increasing estimates of M31's mass.

By matching the weightings to each simulated halo's pericentre time, we obtain our probability distributions for first pericentre time, as illustrated in the right panels of Fig.~\ref{fig:examples}. The upper half of each uses the weightings to select at random an example 100 simulated orbits compatible with the observed satellite, and plots their comoving separation from their hosts through time. The first pericentre time of each can be seen as the first `V' shape below the black dotted line (the haloes' hosts' median virial radii at that time) and these indicate the distribution of the sample's pericentres with time. The lower panel of each is a histogram representation of the full weighted distribution of pericentre times, and serves as the final probability distribution for this quantity.

The orbits shown in the right panel cluster together at lookback time $0\,\mathrm{Gyr}$ because the orbits are constrained to pass through the current uncertain position of the satellite (more so for Andromeda~XIX due to the smaller uncertainty on its distance) and diverge towards the past and the future. There are, however, trends preserved in their histories, as shown in the histograms. For both satellites the distribution forms two peaks, which we here separate for illustrative purposes into a blue peak with old pericentres and a green peak with more recent/future pericentres. This colour split is carried over to the 100 orbital paths, where we see that the green paths are approaching for the first time while blue paths are returning to close radii for a second or even third pericentric passage. This bimodal distribution is seen to some extent in 36 of the 39 satellites, with 5 of these including a third peak, and predominantly originates from a degeneracy due to satellites having broadly similar coordinates at a given phase of the second and each subsequent orbital passage.

These histograms represent our final probability distributions for the pericentre times of Andromedas XVIII and XIX. From these and the distributions for the remaining 37 satellites, we take the $16^\mathrm{th}$, $50^\mathrm{th}$, and $84^\mathrm{th}$ percentiles as our central values and uncertainty intervals. These are imperfect summary statistics for often multi-modal distributions but have the advantage of being concise; we also explore trends that do not use these summary statistics below.

\section{Results and Discussion}
\label{results}

\subsection{M31's satellite population}
\label{mainresults}

\renewcommand{\arraystretch}{1.3}
\begin{table*}
\centering
\caption{Our predictions for orbital parameters, shown as medians with $16^\mathrm{th}$ to $84^\mathrm{th}$ percentile uncertainties for the 39 galaxies studied in this paper, ordered by proximity to M31. Columns: (1) The abbreviated names of the satellite galaxies; (2) the infall times (negative values indicate future events); (3) the pericentre times (negative values again indicate future events); (4) the observed velocity offset from M31 along the line of sight (with the positive direction being away from M31); (5) the predicted magnitude of the velocity component in the plane of the sky and radial to M31; (6) the predicted magnitude of the remaining velocity component in the plane of the sky and perpendicular to the radial direction; (7) the predicted resultant magnitude of the satellites' relative velocities in the plane of the sky ($v_\mathrm{transverse}=(v_\rho^2+v_\phi^2)^{1/2}$); (8) the predicted total magnitude of the satellite's proper velocity relative to M31 ($v_\mathrm{total}=(v_\mathrm{LoS}^2+v_\mathrm{transverse}^2)^{1/2}$). The equations relating components are applied to individual haloes sampled from the simulations and therefore do not necessarily hold for the median values.}
\label{tab:predict}
\begin{tabular}{l|c|c|c|c|c|c|c}
\hline
Satellite galaxy & $t_\mathrm{infall}$ [Gyr] & $t_\mathrm{peri}$ [Gyr] & $v_\mathrm{LoS}$ [km s$^{-1}$] & $|v_\mathrm{\rho}|$ [km s$^{-1}$] & $v_\mathrm{\phi}$ [km s$^{-1}$] & $v_\mathrm{transverse}$ [km s$^{-1}$] & $v_\mathrm{total}$ [km s$^{-1}$] \\
\hline
\input{PredictionsTable.txt} \\
\hline
\end{tabular}
\end{table*}

We first present the median values and uncertainties from our distributions for each satellite in Table~\ref{tab:predict}, ordered by current distance from M31. We show predictions for the satellites' orbital parameters (used in Sec.~\ref{quen} below), predictions of proper motions (which we discuss in Sec.~\ref{prop}), and the observed line of sight velocity offsets with our propagated uncertainties to provide context.

\subsubsection{Orbital histories and quenching}
\label{quen}

To constrain the satellites' orbital histories, we focus on our predicted lookback times to infall and first pericentre. We see from Table~\ref{tab:predict} a large variety of times for these events. There is a trend of more recent infall and pericentre times for satellites currently further from M31 (moving down the table) due to having had less time to settle into a close orbit. There is, however, significant scatter in this trend, for example we find very recent times for Andromeda~XII when compared to satellites at similar distances. This is likely caused by its line-of-sight velocity; Andromeda~XII is currently moving away from M31 at high speed despite being close to M31, suggesting that it has just passed its first pericentre (other galaxies at similar distances are more likely on subsequent orbital passages). The sign of the line-of-sight velocity also provides some otherwise unavailable, valuable information regarding the orbital history, seen by the difference in e.g. NGC~147 and Andromeda~V which have similar distances and line-of-sight speed offsets from M31, but opposite signs to their velocities that translate into different infall and pericentre times. The dependence on mass is discussed below.

\begin{figure}
	\includegraphics[width=\columnwidth]{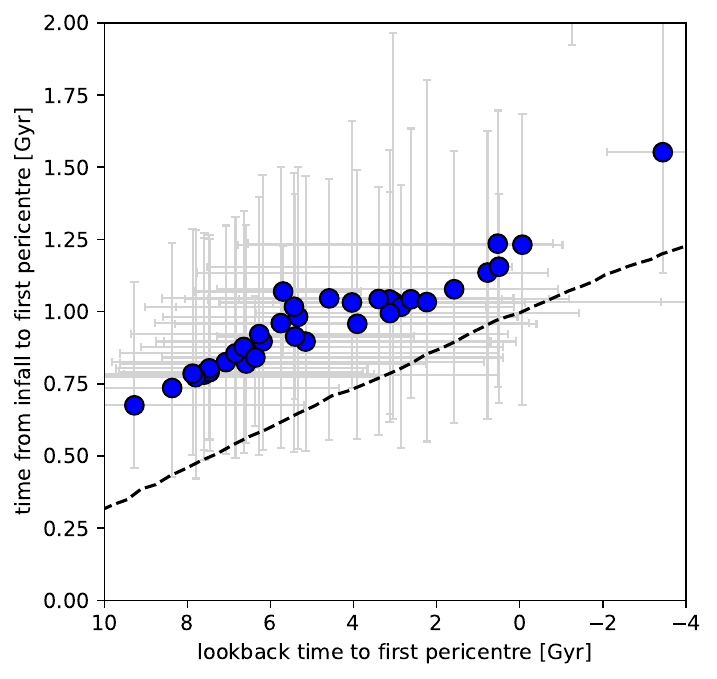}
    \caption{The time we predict to be taken from the first crossing of M31's virial radius to the first pericentric passage for each observed satellite, plotted against the time of its first pericentric passage. This interval is calculated as a probability distribution in its own right, not from our median times for the two events. The black dotted line shows the evolution of the host haloes' median dynamical timescale. The Triangulum galaxy lies outside of the plot area near $(-1,4)\,\mathrm{Gyr}$.}
    \label{fig:inftoperi}
\end{figure}

Infall time and pericentre time are the two major events whose timings we have determined for each satellite. However, the former inevitably leads to the latter. Fig.~\ref{fig:inftoperi} shows the tight correlation between these two times. The time taken from infall to pericentre tends to be around $1\,\mathrm{Gyr}$, with a scatter of about $0.5\,\mathrm{Gyr}$. Pericentric passages occur longer after infall in more recently accreted satellites. This ranges from a median interval of $0.7\,\mathrm{Gyr}$ at $9\,\mathrm{Gyr}$ ago to $1.2\,\mathrm{Gyr}$ at present day. As such there is a simple and shallow linear relation between infall times and pericentre times, reflecting the characteristic dynamical time of M31 through cosmic history. In particular, we show half the crossing time as the black dotted line and calculated as $\frac{1}{2}\sqrt{\frac{r_\mathrm{vir}^3}{\mathrm{G}M_\mathrm{vir}}}$ -- this is expected to be approximately the time taken to reach the galaxy centre after infall. The slope matches that of the trend seen here, despite a $\sim0.3\,\mathrm{Gyr}$ offset, tying the time between infall and pericentre to the host halo's growth. The exception to this relation is Triangulum, landing outside the axes above the plot, which we expect to have almost $4\,\mathrm{Gyr}$ from infall to first pericentre. However, due to its high mass, Triangulum has the fewest matching simulated haloes of our sample -- just 10 -- and this high value is driven by two unusual simulated satellite haloes.  We note that except for 2 cases (Triangulum, and IC~1613 which are amongst the most massive satellites) the rest of the satellites have likely gone through their pericentre in the past or are around their first pericentre now; very recent infalls are unlikely.

We will use pericentre time in the following due to its unambiguous definition (whereas infall time is defined as the crossing time through various radii in the literature), but Fig.~\ref{fig:inftoperi} shows that our results would not be qualitatively altered by using the infall time instead.

\begin{figure}
	\includegraphics[width=\columnwidth]{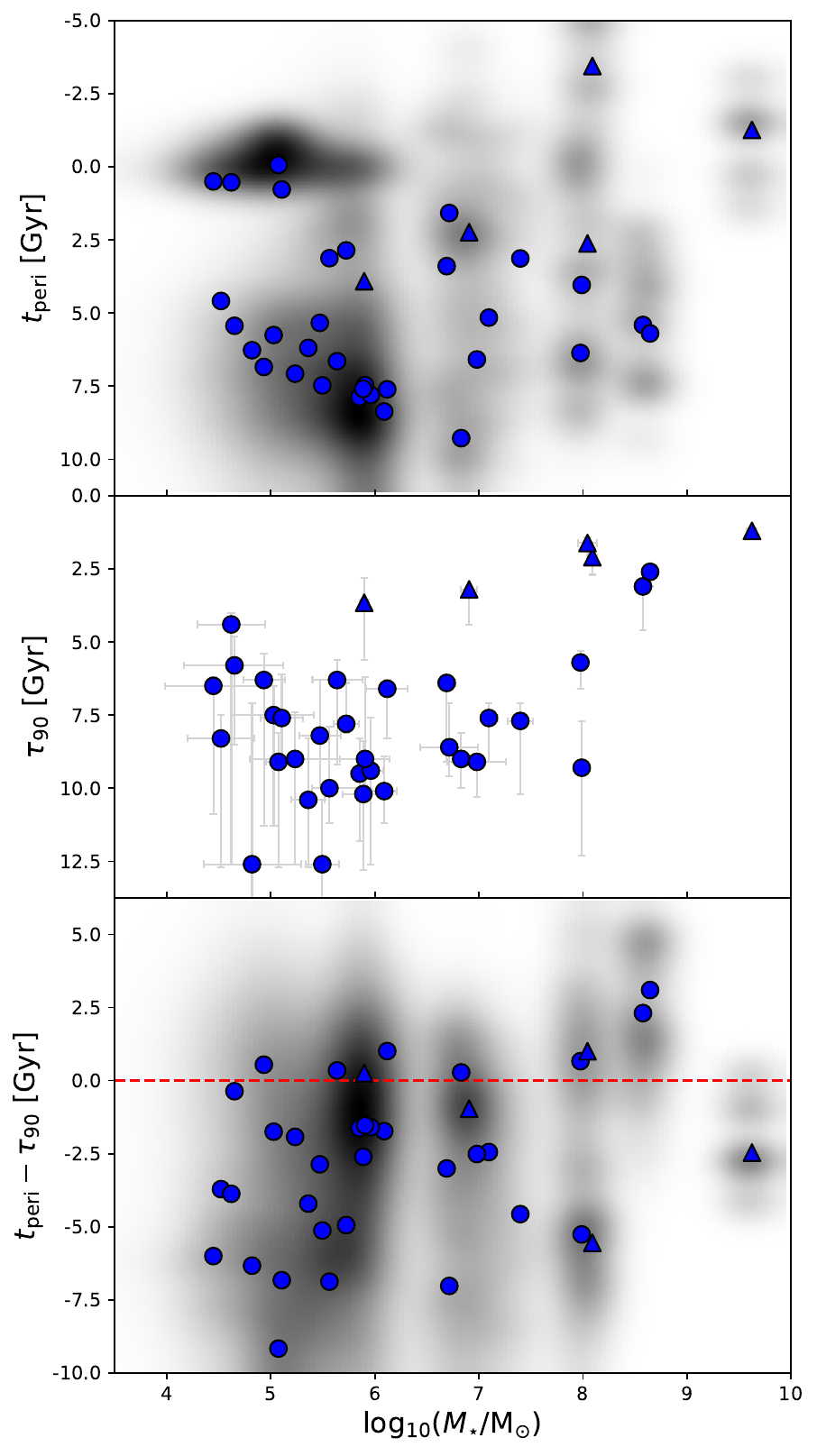}
    \caption{From top to bottom, the blue points show the medians of lookback time to first pericentre passage, lookback time to quenching, and their difference (the quenching timescale), each plotted against the satellites' stellar mass. Galaxies that have not yet quenched (as determined by a morphological classification of dwarf irregular or spiral) are shown by triangles. In the quenching time panel, error bars are used to indicate the $16^\mathrm{th}$ to $84^\mathrm{th}$ percentiles; the top and bottom panels uncertainty is visualised as a summed probability density, with darker regions having a higher probability of hosting an M31 satellite. This is obtained from the values' probability distributions and smoothed using Gaussian kernels across a scale of $0.1\,\mathrm{dex}$ for the logarithmic stellar mass and $0.5\,\mathrm{Gyr}$ for the lookback time ($\sim2$~per~cent of each axis). The dotted line in the bottom panel indicates simultaneous pericentre and quenching times.}
    \label{fig:timescale}
\end{figure}

Fig.~\ref{fig:timescale} shows the dependence of first pericentre time and quenching time (in the top and middle panels, respectively) on the satellites' stellar mass. In the bottom panel we illustrate the relation between stellar mass and quenching timescale, defined as the difference between first perientre time and quenching time (with negative values indicating quenching before first pericentre). The blue points give the median values for each satellite, and the blurred grey scale shows the sum of the probability distributions calculated for all of M31's satellites. For the quenching timescale, we convert the quenching time to a distribution as described in Sec.~\ref{obs} and then combine this with the pericentre time distributions, assuming no covariance between the uncertainties on the two times. The result is a globally summed probability distribution for the position of an M31 satellite within this space of mass and quenching timescale. 

The pericentre time in the top panel forms two tracks at masses below $10^{6.5}\,\mathrm{M}_{\odot}$, one around present day and the other from $4\,\mathrm{Gyr}$ to $8\,\mathrm{Gyr}$ ago from low to high mass, before combining into a wider scatter at higher masses. This is a result of the prevalent bi-modality of the probability distributions for orbital history parameters across most of the satellites at low masses (see Sec.~\ref{distributions}), creating a clear separation of pericentre times once combined into an overall distribution. Additionally, the median values along the early-pericentre track are mainly composed of satellites which are currently closer to M31, while the more recent track consists of more distant satellites.

In the quenching times, taken from the literature and shown in the second panel, there is a trend of later quenching times with increasing mass which is well established: higher mass dwarfs keep forming stars for most of the cosmic time whereas lower mass objects are on average quenched a long time ago. The quenching timescales shown in the bottom panel have a trend mostly inherited from the trend in the quenching times (middle panel), with higher mass galaxies generally having less negative timescales, albeit with a large amount of scatter. In fact, only 2 of the most massive satellites (M32 and NGC~205), have a high probability of having quenched more than $1\,\mathrm{Gyr}$ after their first pericentre passage, while the 5 that are yet to quench, shown by the triangles, tend towards higher masses as well. Ultimately, we find that of the 7 most massive satellites with $M_\star>10^8\,\mathrm{M}_\odot$, only one is likely to have been quenched before first pericentre (NGC~185) - the other 6 have a positive quenching timescale and/or have not yet quenched from morphological classification. In contrast, the majority of the remaining lower mass satellites appear to have been quenched significantly before their first pericentre passage, with some of the least massive quenching up to $10\,\mathrm{Gyr}$ prior.

%The majority of satellites with stellar masses below $10^{8}\,\mathrm{M}_{\odot}$ probably quench before their first pericentre passage. Lower mass galaxies are the most vulnerable to environmental quenching mechanisms, and this cut-off would suggest that they are unable to maintain star formation through their infall onto M31. Therefore, for those low-mass satellites quenching less than $\sim1\,\mathrm{Gyr}$ before their first pericentre, ram-pressure stripping seems like the best candidate to be the dominant quenching mechanism. Some of the low mass satellites which quench well before encountering M31 may have encountered other high density environments before M31 and been quenched by cosmic web stripping or, for the earliest quenching galaxies in our sample, by cosmic reionization.

%For stellar masses from $10^{5.5}\,\mathrm{M}_{\odot}$ to $10^{7.5}\,\mathrm{M}_{\odot}$, there are both positive and negative quenching timescales, however none significantly greater than $2.5\,\mathrm{Gyr}$. This indicates that star formation in satellites in this mass range, while able to survive infall onto M31, is unable to survive the ram-pressure and tidal effects of their first pericentres, with all being quenched by around the time of their first apocentre. Again, some of these satellites appear to have quenched before interacting with M31, particularly towards the lower mass end of this group.

In the background distribution (greyscale), we still see the general trend of more positive quenching timescales with higher mass. This is particularly prominent in the over-density at $-6\,\mathrm{Gyr}$ below $10^{5.5}\,\mathrm{M}_{\odot}$, which sharply rises to around $1\,\mathrm{Gyr}$ by $10^6\,\mathrm{M}_{\odot}$. The higher values of quenching timescale are maintained by the over-densities at higher masses, although not without significant scatter. Large uncertainties make the trends previously described weaker, but it is still clear that the lowest-mass satellites do not maintain star formation long past their first pericentre passage, whereas the most massive satellites have the potential not to quench until long after this event.

This all suggests that only the most massive satellites can resist environmental quenching from the environment surrounding M31 for any significant length of time. Below a stellar mass of $\sim10^8\,\mathrm{M}_\odot$, a number of satellites have quenching timescales consistent with being quenched by tidal stripping and ram-pressure stripping effects near their first pericentre around M31. However, the majority of M31's satellite population may have quenched well before interacting with M31. This could hint at ram-pressure stripping when passing through filaments of the cosmic web \citep{2013Be}, or `pre-processing' around lower mass hosts. Additionally, a few of the satellites quenched earliest may have been quenched by reionization; Andromeda~XI and Andromeda~XVII are strong candidates for this mechanism, both being at the lower mass end of our sample and quenching around $12.5\,\mathrm{Gyr}$ ago in spite of their likely infall times being only $7$--$8\,\mathrm{Gyr}$ ago.

%This all suggests that only the satellites with stellar masses greater than $10^{7.5}\,\mathrm{M}_\odot$ are more resistant to the quenching mechanisms from the environment surrounding M31 and are able to hold on to their cold gas for longer. Below this mass, most satellites' histories appear consistent with having been quenched by tidal stripping and ram-pressure stripping effects from their dense environment shortly after their first pericentre passage. A smaller number quenched just before pericentre, suggesting that ram-pressure stripping on infall could be the dominant mechanism for quenching in these satellites. While large uncertainties in the pericentre times for each satellite mean the relative importance of these two stripping mechanisms is unclear, their combination appears to reliably quench these faint dwarf satellites. Lastly, there are a number of satellites which likely quenched well before interacting with M31. $\sim4$ have quenching times consistent with the epoch of reionization, while the remainder likely underwent some prior interaction which quenched them early in their lives.

\subsubsection{Predictions of proper motions}
\label{prop}

Ever increasing observational capabilities will likely yield proper motions for many of M31's satellites this decade, so we predict these using our methodology to act as a future validation test of the accuracy of our predictions of significant orbital events. These proper motions relative to M31 are shown in Table~\ref{tab:predict}. Many of these velocity predictions have very large uncertainties. Some, however, are more precise, driven by orbits that require a velocity in a narrow range at the present day. Note that, while we have signed $v_\rho$ available from our data, we show the absolute value $|v_\rho|$ since many of the satellites give distributions roughly symmetric about $0\,\mathrm{km}\,\mathrm{s}^{-1}$.

\begin{figure}
	\includegraphics[width=\columnwidth]{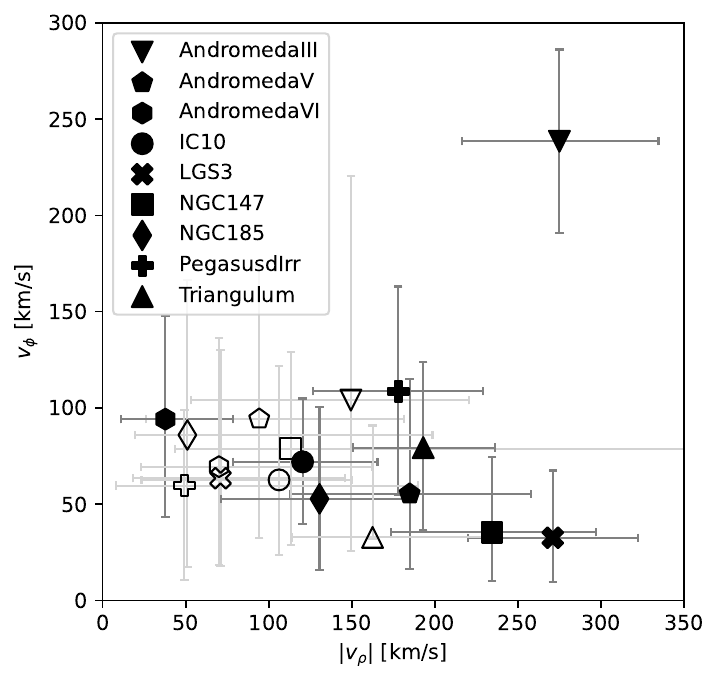}
    \caption{The proper motions in the plane of the sky relative to M31 for the 9 satellites with observed proper motions available. Filled points show these observed values, while open points show our predictions from Table~\ref{tab:predict}.}
    \label{fig:proper}
\end{figure}

%\begin{table*}
%    \centering
%    \caption{The space motions in the plane of the sky relative to M31 for the 5 satellites with measured proper motions. Columns: (1) The abbreviated names of the satellite galaxies; (2) the magnitude of the velocity component in the plane of the sky and radial to M31 (with the positive direction being radially outwards from M31); (3) the magnitude of the remaining velocity component in the plane of the sky and perpendicular to the radial direction; (4) the resultant magnitude of the satellites' relative velocities in the plane of the sky ($v_\mathrm{transverse}=(v_\rho^2+v_\phi^2)^{1/2}$); (5) the references from which these velocities are calculated. (a) \citet{2007Br} (b) \citet{2020So} (c) \citet{2024CD} (d) \citet{2024Ru}}
%    \begin{tabular}{l|c|c|c|c}
%        \hline
%        Satellite & $v_\mathrm{\rho}$ [km s$^{-1}$] & $v_\mathrm{\phi}$ [km s$^{-1}$] & $v_\mathrm{transverse}$ [km s$^{-1}$] & References \\
%        \hline
%        Triangulum & $-194_{-43}^{+44}$ & $79_{-42}^{+44}$ & $213_{-46}^{+47}$ & (d) \\
%        IC~10 & $122_{-43}^{+44}$ & $73_{-32}^{+33}$ & $147_{-37}^{+39}$ & (a) (d) \\
%        NGC~147 & $234_{-61}^{+61}$ & $37_{-25}^{+40}$ & $241_{-60}^{+60}$ & (b) (d) \\
%        NGC~185 & $131_{-58}^{+60}$ & $54_{-37}^{+47}$ & $149_{-53}^{+56}$ & (b) (d) \\
%        And~III & $-276_{-57}^{+60}$ & $239_{-46}^{+47}$ & $367_{-56}^{+58}$ & (c) (d) \\
%        \hline
%    \end{tabular}
%    \label{tab:properobs}
%\end{table*}

Measurements are available for the proper motions of 9 of M31's satellites - Andromeda~III \citep{2024CD}, Andromeda~V and Andromeda~VI \citep{2025CD}, IC~10 \citep{2007Br}, NGC~147 and NGC~185 \citep{2020So}, LGS~3 and PegasusdIrr \citep{2025Be}, and Triangulum \citep{2024Ru}. With the addition of M31 itself \citep{2024Ru}, this allows us to compare these with the values predicted using our methodology. To compare against our predictions, we convert each proper motion to components matching those in Table~\ref{tab:predict}, since these components are independent of orientation in the plane of the sky which we do not constrain in our simulated halo selection. We show these velocity components in Fig.~\ref{fig:proper} alongside our predictions. The $\sim 68$~per~cent confidence regions for the predictions and measurements for Andromeda~V, Andromeda~VI, IC~10, NGC~185, PegasusdIrr and Triangulum (6/9 galaxies; 67~per~cent) overlap, while predictions and measurements for Andromeda~III and NGC~147 are tentatively compatible with overlapping $\sim 95$~per~cent confidence regions (not shown). Our predictions are incompatible with observations for LGS~3. Across these 9 satellites, our predictions therefore seem to give statistically reliable results, albeit with large uncertainties.

% We see that of these 5, only IC~10 produces a good match between prediction and observation, while Andromeda~III instead sees a definite discrepancy. The remaining three satellites each have predictions lying within observational uncertainties in one of their two components. For these three, it should be noted that in their less well matching component the predicted uncertainties are large enough to include the central observed values. Overall, this suggests that our methodology produces somewhat accurate but imprecise predictions of proper motions.

For Sec.~\ref{quen}, the comparison with observed proper motion measurements suggests that median values carry uncertainties too large to be reliable indicators of satellites' individual orbital histories. However, by taking these as a population and using all of the information in our probability distributions, we expect that we have recovered a reasonably accurate description of M31's overall satellite population.

\subsection{Milky~Way comparison}
\label{mw-comparison}

As mentioned in Sec.~\ref{intro}, the precision of measurements in the Milky~Way and its satellite system have enabled detailed study to precede that in external galaxies. \citet{2019Fi} performed a similar study to the one that we present here, but for the Milky~Way's satellites, using 12 simulated haloes and observations constraining the full 6D phase space. In this section, we use this to compare our results for M31's satellites with the infall and quenching of the Milky~Way's satellites.

\begin{figure}
	\includegraphics[width=\columnwidth]{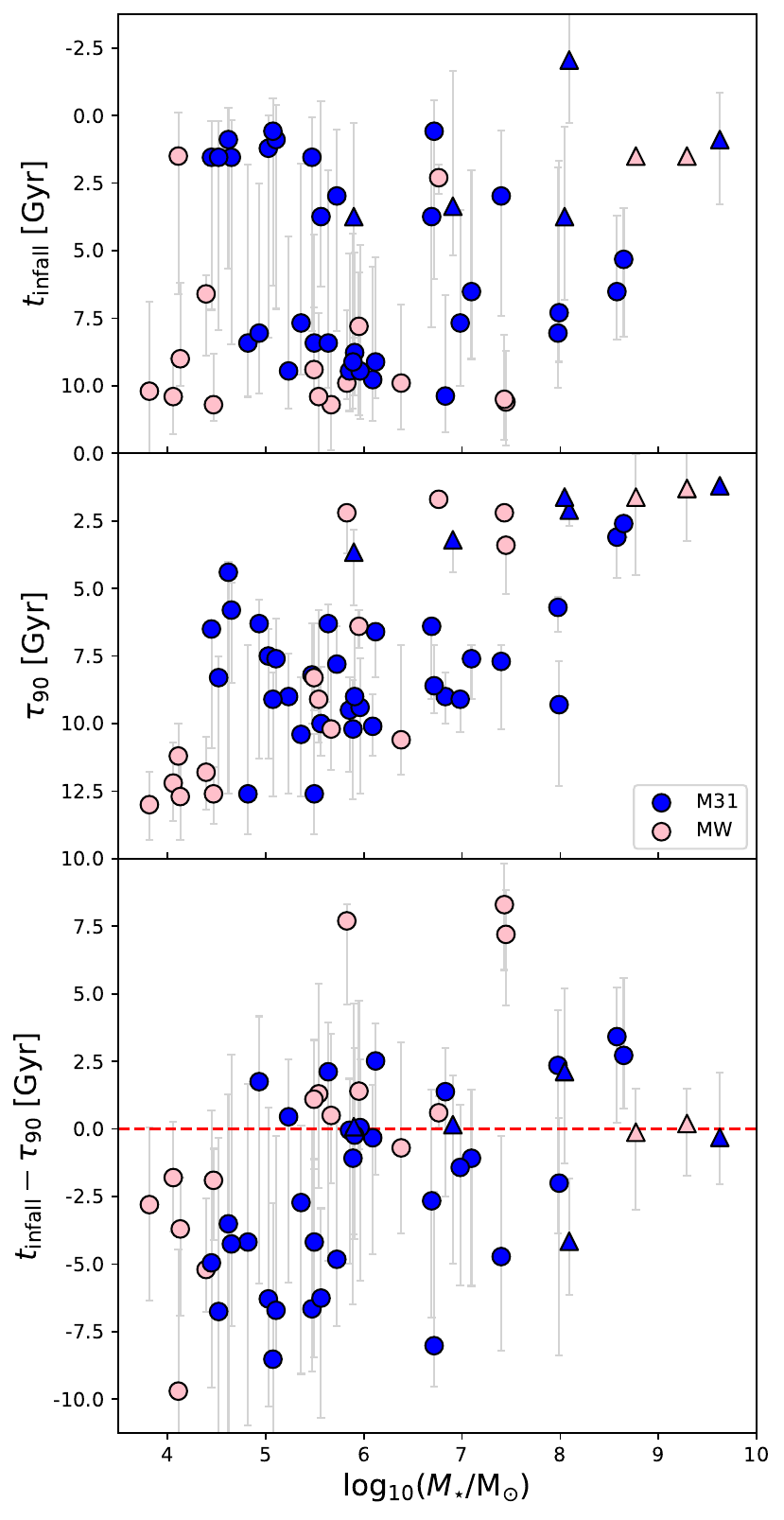}
    \caption{From top to bottom, the lookback time to infall, the lookback time to quenching, and their difference (the infall-quenching timescale) are plotted against stellar mass. Uncertainties in time are shown by the error bars in each panel. Milky~Way satellites are shown in light pink, while M31 satellites are shown in blue. Satellites that are yet to quench are shown as triangles, with the same meaning for their colours. The dotted line in the bottom panel signifies equal times of quenching and infall, with points below this line being quenched before infall.}
    \label{fig:mwcomp}
\end{figure}

Fig.~\ref{fig:mwcomp} compares the trends in infall time, quenching time and quenching timescale with mass of satellites of M31 against those of the Milky~Way. We obtain the Milky~Way satellites' stellar masses from tables from \citet{2012MC} using the same method as described in Sec.~\ref{obsprocessing}. For most of the Milky~Way satellites, we take infall and quenching times from \citet{2019Fi}, but for the Small Magellanic Cloud (SMC) and Large Magellanic Cloud (LMC) we use infall times (without uncertainties) from the most likely orbital history found in \citet{2013Ka} and values of $\tau_{90}$ from \citet{2018Ru} and \citet{2021Ma}. For a direct comparison with \citet{2019Fi}, we use infall times rather than pericentre times in this section, although we again argue using Fig.~\ref{fig:inftoperi} that these two measures produce qualitatively equivalent results. We also replace our previously used median with the mode of the smoothed histograms of the probability distributions for infall times to better mimic the approach of \citet{2019Fi}. Due to the prominence of bimodal distributions in these probabilities, this generally has the effect of shifting the estimated infall time towards more extreme values, either close to present day or close to lookback times of $9-10\,\mathrm{Gyr}$, depending on which peak in the probability distribution has the largest amplitude.

Despite this, we still see a more continuous distribution of infall times for satellites of M31 than for those around the Milky~Way in the top panel of Fig.~\ref{fig:mwcomp}, with the same being seen for quenching times in the second panel. As \citet{2019Fi} notes, the Milky~Way's satellites mostly have old ($>11\,\mathrm{Gyr}$) quenching times and/or old ($>9\,\mathrm{Gyr}$) infall times, accounting for $76$~per~cent of the population. In contrast, M31's satellites are more widely and evenly spread in both infall and quenching times.

This suggests some systematic difference between the two satellite populations, either from observational biases or genuine differences. \citet{2019Fi} use a set of zoom-in simulations to infer infall times, and have proper motions available for their halo-matching. As such, they have better observational constraints on the Milky~Way satellites' orbits (thanks to the proper motions), and improved mass resolution in their simulations (probably necessary to track the orbits of the halos hosting ultra-faint satellites). However, this comes at the cost of a smaller sample of hosts and satellites than in this study, which has a greater variety of satellite histories to consider as a result. \citet{2019Fi} also reach lower mass satellites than for M31, down to $M_\star=10^{3.5}\,\mathrm{M}_\odot$. These galaxies are more likely to be quenched by reionization; they make up the extremely early quenched group seen in the lower left of Fig.~\ref{fig:mwcomp}'s second panel.

Assuming that the differences are instead physical, and ignoring those Milky~Way satellites below the mass range for M31's observed satellites, the primary remaining difference is a scarcity of recently infalling satellites around the Milky~Way. This would suggest that the Milky~Way generally assembled its satellites earlier than M31, with LMC and SMC being notable recent exceptions. Indeed, this conclusion is also reached following different lines of argument in other studies based on either observations \citep[e.g.][]{2007Ha,2019Kr,2020Fo} or simulations \citep[e.g.][]{2020Fr,2020Ev,2024Bu,2025We}. The consensus emerging is that the Milky~Way has assembled its mass and satellite population unusually early. M31 seems to be a more typical galaxy in this regard. Additionally, \citet{2017Ge,2024Ge} find that the Milky~Way's satellites have a higher quenched fraction than typical for similar hosts in the local Universe. \citet{2023Gr} also find that other satellite systems see a less abrupt change with mass from fast to slow quenching times, which matches the more varied distribution of quenching times for M31's satellites that we see here and was also seen in \citet{2025Sa}.

In the bottom panel of Fig.~\ref{fig:mwcomp}, we show an altered version of our quenching timescale, with pericentre time replaced by infall time. We can see that while very few Milky~Way satellites maintain star formation past $2\,\mathrm{Gyr}$ after infall below $10^8\,\mathrm{M}_\odot$, there are three that have have probably survived for at least $7\,\mathrm{Gyr}$. \citet{2019Fi} identify all three as unusual; the Sagittarius dSph is being extremely tidally stripped which may lead to its genuine counterparts in simulations being interpreted as destroyed, while Fornax and Carina are both on extremely circular orbits with distant pericentres and so subject to milder ram-pressure and/or tidal stripping from the Milky~Way. This leaves the sample of more typical Milky~Way satellites in two groups: those which were quenched around infall, with  stellar masses above $10^{5}\,\mathrm{M}_\odot$, and those which were quenched before infall, with stellar masses lower than this. The higher mass group was likely quenched by ram-pressure/tidal stripping from the Milky~Way, while the lower mass group are those consistent with quenching by reionisation.

The largest difference between the two satellite populations is the majority of M31's satellites across all except the highest masses which can be seen to quench before infall, as in Fig.~\ref{fig:timescale}. This population of negative quenching timescales is not present above $10^{5}\,\mathrm{M}_\odot$ in the Milky~Way's satellites, possibly due to the overwhelmingly early infall times in the Milky~Way. M31's more varied distribution of satellite infall times may uncover the importance of pre-processing quenching where the Milky~Way's unusual satellite population hides this.

\subsection{Quenching mechanisms in context}
\label{literature}

Other recent studies have also placed constraints on the mechanisms driving quenching in satellite galaxies. In this section, we compare our results with those found in other works to place our conclusions in context.

Ram-pressure stripping has been seen in external satellite systems \citep{2024Jo} and is consistent with having quenched $80$~per~cent of the Milky~Way's satellite population \citep{2021Pu}. In a cosmological simulation, \citet{2023En} find that ram-pressure stripping in combination with tidal stripping is the predominant cause of quenching for satellite galaxies with stellar masses below $\sim10^9\,\mathrm{M}_\odot$ around M31-mass hosts. Similarly, \citet{2021DC} find that ram-pressure stripping quickly quenches star formation in $75$~per~cent of the satellites in a simulated Local Group analogue system. While our methodology cannot reliably differentiate between ram-pressure and tidal stripping mechanisms, we do see a significant number of M31's satellites quenching around their first pericentres, with a rough upper limit on the quenching timescale of $2.5\,\mathrm{Gyr}$ for satellites with stellar masses less than $\sim10^8\,\mathrm{M}_\odot$. This timescale is consistent with both gas stripping mechanisms, and additionally agrees with the $3\,\mathrm{Gyr}$ after infall quenching timescale found in \citet{2020Ak} for the same mass limit from zoom simulations.

However, other works find vastly different quenching timescales. \citet{2025Pa} find timescales of $\sim300\,\mathrm{Myr}$ for simulated satellites infalling over $10\,\mathrm{Gyr}$ ago and \citet{2023Ba} find only slightly longer timescales of $\sim0.5\,\mathrm{Gyr}$ around the Milky~Way. In contrast, \citet{2023Gr} find timescales $3\,\mathrm{Gyr}$ or less only for observed satellites with stellar masses below around $10^6\,\mathrm{M}_\odot$, and \citet{2021Om} find timescales on the order of the Hubble time for $M_\star\sim10^9\,\mathrm{M}_\odot$ satellites around slightly more massive hosts. It appears that quenching timescales may not be consistent across all systems, but ram-pressure and tidal stripping are likely important mechanisms of quenching in many, including in M31's satellites.
%Such results conflict with ram-pressure and tidal stripping as the ubiquitous, dominant mechanisms of satellite quenching, but there is still abundant evidence to suggest that it is an important process.

Across M31's satellites, we see most quenching occuring before pericentre or even before infall. This is consistent with any of internal quenching, pre-processing by other halos and/or the cosmic web or, in a few of the oldest cases, quenching by reionisation. A reliance on pre-processing was also found in zoom simulations by \citet{2025Jo}, where many satellites below $10^8\,\mathrm{M}_\odot$ stellar mass around Milky~Way-mass galaxies quenched outside the host environment over $8\,\mathrm{Gyr}$ ago (these results differ depending on the parent simulation used). Pre-processing from the cosmic web has also recently been seen in observations in \citet{2025Lu}. Closer to this study, in the Local Group, \citet{2025Be} find pre-processing to be an important factor in the quenching of the outermost satellites. Reionisation in the Milky~Way system on the other hand, is only expected to be relevant for satellites below $10^5\,\mathrm{M}_\odot$ stellar mass \citep{2023Ba}. This excludes most of our sample of M31 satellites, with only a handful of good candidates for reionisation, leaving pre-processing as the most likely mechanism of quenching for the majority of M31's satellites population.

\section{Conclusions}
\label{conclusions}

In this work we used the M31 system to investigate the mechanisms that drive the quenching of satellite galaxies, and presented predictions for the proper motions of satellites of M31. We constructed probability distributions of each satellite's observed kinematic properties and used these to select and weight matching satellite haloes from a dark matter only simulation. From these haloes' orbital histories, we assigned each of M31's satellites a probability distribution for their proper motions, time of infall and first pericentre around M31.

\begin{itemize}
  \item We see that only the most massive of M31's satellites have maintained star formation for more than $3\,\mathrm{Gyr}$ following their first pericentre. This suggests that ram-pressure, tidal stripping and/or the shutoff of gas accretion reliably quenches dwarf galaxies with $M_\star<10^{7.5}\,\mathrm{M}_\odot$ upon becoming satellites of M31.
  \item A significant portion of M31's lower mass satellites likely quenched well before encountering M31, some possibly from reionization but most from 'pre-processing'.
  \item We present predictions for 39 satellites' proper motions, to be compared with future observations (9 galaxies with existing proper motion measurements are consistent with our predictions, although the uncertainties are large).
  \item By comparing with the Milky~Way, using data from \citet{2019Fi}, we find that the two galaxies have qualitatively different satellite populations. In particular, the Milky~Way's satellites have generally been satellites for longer and have been quenched more quickly following infall than M31's satellite population.
\end{itemize}

With probable orbital histories now available for satellite galaxies across the Local Group, we have a wider sample than ever for examining how satellites interact with their hosts. The properties of M31's satellites reflect the fact that environmental effects -- ram-pressure, tidal stripping or the cessation of gas accretion -- are reliable quenchers of low mass satellite galaxies in the Universe.

\section*{Acknowledgements}

KAO acknowledges support by the Royal Society through Dorothy Hodgkin Fellowship DHR/R1/231105, by STFC through grant ST/T000244/1, and by the European Research Council (ERC) through an Advanced Investigator Grant to C.~S. Frenk, DMIDAS (GA 786910). AF acknowledges support by a UKRI Future Leaders Fellowship (grant no MR/T042362/1) and a Wallenberg Academy Fellowship. This work used the DiRAC@Durham facility managed by the Institute for Computational Cosmology on behalf of the STFC DiRAC HPC Facility (\url{www.dirac.ac.uk}). The equipment was funded by BEIS capital funding via STFC capital grants ST/K00042X/1, ST/P002293/1, ST/R002371/1 and ST/S002502/1, Durham University and STFC operations grant ST/R000832/1. DiRAC is part of the National e-Infrastructure. This research has made use of NASA's Astrophysics Data System.

%%%%%%%%%%%%%%%%%%%%%%%%%%%%%%%%%%%%%%%%%%%%%%%%%%
\section*{Data Availability}

The data presented in this article and the underlying data distributions will be will be available at https://icc.dur.ac.uk/data/ on publication.



\bibliographystyle{mnras}
\bibliography{example} % if your bibtex file is called example.bib

%\begin{thebibliography}{99}
%\end{thebibliography}

%%%%%%%%%%%%%%%%%%%%%%%%%%%%%%%%%%%%%%%%%%%%%%%%%%

% Don't change these lines
\bsp	% typesetting comment
\label{lastpage}
\end{document}

%% file: PredictionsTable.txt
M32 & $ 6.3 _{- 2.1 } ^{+ 1.9 } $ & $ 5.4 _{- 2.9 } ^{+ 1.9 } $ & $ -88 _{- 21 } ^{+ 193 } $ & $ 61 _{- 16 } ^{+ 129 } $ & $ 36 _{- 13 } ^{+ 178 } $ & $ 114 _{- 59 } ^{+ 223 } $ & $ 162 _{- 53 } ^{+ 186 } $ \\
And~I & $ 9.7 _{- 3.5 } ^{+ 1.5 } $ & $ 9.3 _{- 4.1 } ^{+ 1.3 } $ & $ 60 _{- 140 } ^{+ 20 } $ & $ 135 _{- 75 } ^{+ 98 } $ & $ 126 _{- 90 } ^{+ 75 } $ & $ 200 _{- 67 } ^{+ 94 } $ & $ 208 _{- 61 } ^{+ 100 } $ \\
And~XVII & $ 8.1 _{- 3.2 } ^{+ 2.2 } $ & $ 7.5 _{- 3.8 } ^{+ 2.3 } $ & $ -45 _{- 8 } ^{+ 93 } $ & $ 170 _{- 113 } ^{+ 133 } $ & $ 193 _{- 106 } ^{+ 119 } $ & $ 307 _{- 107 } ^{+ 124 } $ & $ 311 _{- 107 } ^{+ 123 } $ \\
NGC~205 & $ 6.7 _{- 2.0 } ^{+ 1.9 } $ & $ 5.7 _{- 1.9 } ^{+ 2.1 } $ & $ 54 _{- 6 } ^{+ 7 } $ & $ 102 _{- 58 } ^{+ 169 } $ & $ 68 _{- 61 } ^{+ 109 } $ & $ 201 _{- 148 } ^{+ 117 } $ & $ 206 _{- 130 } ^{+ 147 } $ \\
And~XXVII & $ 6.6 _{- 5.4 } ^{+ 3.2 } $ & $ 6.0 _{- 6.0 } ^{+ 3.2 } $ & $ -235 _{- 26 } ^{+ 30 } $ & $ 87 _{- 57 } ^{+ 92 } $ & $ 86 _{- 58 } ^{+ 95 } $ & $ 148 _{- 67 } ^{+ 99 } $ & $ 301 _{- 76 } ^{+ 85 } $ \\
And~IX & $ 7.7 _{- 5.3 } ^{+ 2.6 } $ & $ 7.1 _{- 6.2 } ^{+ 2.8 } $ & $ -91 _{- 10 } ^{+ 8 } $ & $ 135 _{- 90 } ^{+ 120 } $ & $ 156 _{- 107 } ^{+ 123 } $ & $ 248 _{- 107 } ^{+ 97 } $ & $ 266 _{- 100 } ^{+ 95 } $ \\
And~III & $ 8.9 _{- 3.4 } ^{+ 2.0 } $ & $ 8.4 _{- 4.0 } ^{+ 2.1 } $ & $ 44 _{- 6 } ^{+ 6 } $ & $ 149 _{- 96 } ^{+ 71 } $ & $ 104 _{- 78 } ^{+ 117 } $ & $ 204 _{- 84 } ^{+ 88 } $ & $ 210 _{- 82 } ^{+ 85 } $ \\
And~XXV & $ 8.2 _{- 6.0 } ^{+ 2.2 } $ & $ 7.6 _{- 7.1 } ^{+ 2.4 } $ & $ -186 _{- 22 } ^{+ 371 } $ & $ 96 _{- 64 } ^{+ 85 } $ & $ 78 _{- 58 } ^{+ 118 } $ & $ 145 _{- 73 } ^{+ 125 } $ & $ 257 _{- 81 } ^{+ 112 } $ \\
And~XV & $ 8.3 _{- 4.5 } ^{+ 2.6 } $ & $ 7.8 _{- 4.4 } ^{+ 2.5 } $ & $ 22 _{- 10 } ^{+ 5 } $ & $ 142 _{- 89 } ^{+ 111 } $ & $ 101 _{- 61 } ^{+ 103 } $ & $ 204 _{- 84 } ^{+ 106 } $ & $ 205 _{- 82 } ^{+ 106 } $ \\
And~XI & $ 7.0 _{- 5.5 } ^{+ 2.9 } $ & $ 6.3 _{- 6.0 } ^{+ 3.1 } $ & $ 116 _{- 229 } ^{+ 15 } $ & $ 118 _{- 78 } ^{+ 118 } $ & $ 139 _{- 93 } ^{+ 98 } $ & $ 216 _{- 83 } ^{+ 87 } $ & $ 252 _{- 78 } ^{+ 87 } $ \\
And~XXVI & $ 7.5 _{- 5.9 } ^{+ 2.6 } $ & $ 6.8 _{- 6.3 } ^{+ 2.8 } $ & $ 34 _{- 74 } ^{+ 9 } $ & $ 138 _{- 85 } ^{+ 106 } $ & $ 152 _{- 98 } ^{+ 99 } $ & $ 231 _{- 72 } ^{+ 94 } $ & $ 236 _{- 70 } ^{+ 93 } $ \\
NGC~147 & $ 7.1 _{- 5.5 } ^{+ 2.3 } $ & $ 6.4 _{- 6.0 } ^{+ 2.3 } $ & $ -106 _{- 12 } ^{+ 10 } $ & $ 113 _{- 70 } ^{+ 286 } $ & $ 79 _{- 50 } ^{+ 50 } $ & $ 180 _{- 75 } ^{+ 175 } $ & $ 218 _{- 65 } ^{+ 161 } $ \\
And~XII & $ 1.6 _{- 0.3 } ^{+ 6.5 } $ & $ 0.5 _{- 0.3 } ^{+ 7.0 } $ & $ 257 _{- 21 } ^{+ 27 } $ & $ 74 _{- 50 } ^{+ 89 } $ & $ 84 _{- 62 } ^{+ 78 } $ & $ 136 _{- 70 } ^{+ 78 } $ & $ 304 _{- 80 } ^{+ 85 } $ \\
And~V & $ 8.6 _{- 4.1 } ^{+ 2.0 } $ & $ 7.9 _{- 4.4 } ^{+ 2.2 } $ & $ 98 _{- 199 } ^{+ 14 } $ & $ 94 _{- 68 } ^{+ 87 } $ & $ 94 _{- 62 } ^{+ 80 } $ & $ 162 _{- 68 } ^{+ 81 } $ & $ 197 _{- 58 } ^{+ 79 } $ \\
And~XIX & $ 7.3 _{- 5.7 } ^{+ 2.2 } $ & $ 6.6 _{- 6.2 } ^{+ 2.5 } $ & $ 185 _{- 29 } ^{+ 21 } $ & $ 86 _{- 60 } ^{+ 97 } $ & $ 74 _{- 49 } ^{+ 91 } $ & $ 151 _{- 78 } ^{+ 81 } $ & $ 254 _{- 73 } ^{+ 80 } $ \\
And~XXI & $ 8.1 _{- 3.1 } ^{+ 2.1 } $ & $ 7.5 _{- 3.7 } ^{+ 2.2 } $ & $ 55 _{- 120 } ^{+ 12 } $ & $ 108 _{- 75 } ^{+ 102 } $ & $ 113 _{- 81 } ^{+ 97 } $ & $ 182 _{- 78 } ^{+ 90 } $ & $ 195 _{- 71 } ^{+ 90 } $ \\
And~XIII & $ 6.2 _{- 4.9 } ^{+ 3.3 } $ & $ 5.4 _{- 5.3 } ^{+ 3.6 } $ & $ 113 _{- 12 } ^{+ 13 } $ & $ 120 _{- 80 } ^{+ 104 } $ & $ 124 _{- 81 } ^{+ 89 } $ & $ 202 _{- 77 } ^{+ 81 } $ & $ 232 _{- 66 } ^{+ 87 } $ \\
And~XX & $ 5.5 _{- 4.2 } ^{+ 3.6 } $ & $ 4.6 _{- 4.4 } ^{+ 4.0 } $ & $ 153 _{- 23 } ^{+ 18 } $ & $ 105 _{- 68 } ^{+ 102 } $ & $ 122 _{- 86 } ^{+ 90 } $ & $ 185 _{- 75 } ^{+ 93 } $ & $ 247 _{- 63 } ^{+ 96 } $ \\
And~XXIII & $ 8.3 _{- 3.3 } ^{+ 1.8 } $ & $ 7.6 _{- 3.8 } ^{+ 1.9 } $ & $ -61 _{- 8 } ^{+ 114 } $ & $ 99 _{- 77 } ^{+ 96 } $ & $ 93 _{- 70 } ^{+ 118 } $ & $ 165 _{- 80 } ^{+ 93 } $ & $ 179 _{- 70 } ^{+ 91 } $ \\
And~XXXII & $ 7.2 _{- 3.4 } ^{+ 2.8 } $ & $ 6.6 _{- 4.3 } ^{+ 3.0 } $ & $ -70 _{- 9 } ^{+ 136 } $ & $ 97 _{- 69 } ^{+ 77 } $ & $ 59 _{- 41 } ^{+ 95 } $ & $ 135 _{- 70 } ^{+ 87 } $ & $ 156 _{- 61 } ^{+ 82 } $ \\
NGC~185 & $ 4.9 _{- 3.6 } ^{+ 3.0 } $ & $ 4.0 _{- 4.2 } ^{+ 3.2 } $ & $ -96 _{- 11 } ^{+ 8 } $ & $ 51 _{- 32 } ^{+ 148 } $ & $ 86 _{- 69 } ^{+ 80 } $ & $ 165 _{- 115 } ^{+ 89 } $ & $ 194 _{- 91 } ^{+ 78 } $ \\
And~X & $ 6.5 _{- 5.4 } ^{+ 2.8 } $ & $ 5.8 _{- 6.0 } ^{+ 3.0 } $ & $ -136 _{- 14 } ^{+ 11 } $ & $ 86 _{- 61 } ^{+ 94 } $ & $ 87 _{- 61 } ^{+ 87 } $ & $ 149 _{- 70 } ^{+ 88 } $ & $ 210 _{- 58 } ^{+ 79 } $ \\
And~XIV & $ 6.2 _{- 4.8 } ^{+ 2.8 } $ & $ 5.3 _{- 5.3 } ^{+ 3.2 } $ & $ 162 _{- 349 } ^{+ 28 } $ & $ 74 _{- 50 } ^{+ 81 } $ & $ 68 _{- 45 } ^{+ 89 } $ & $ 122 _{- 57 } ^{+ 84 } $ & $ 237 _{- 65 } ^{+ 76 } $ \\
And~II & $ 5.8 _{- 3.8 } ^{+ 3.1 } $ & $ 5.2 _{- 4.5 } ^{+ 3.4 } $ & $ -108 _{- 12 } ^{+ 9 } $ & $ 84 _{- 52 } ^{+ 87 } $ & $ 75 _{- 54 } ^{+ 62 } $ & $ 133 _{- 63 } ^{+ 87 } $ & $ 179 _{- 56 } ^{+ 66 } $ \\
And~XXIX & $ 6.9 _{- 5.4 } ^{+ 2.4 } $ & $ 6.2 _{- 6.1 } ^{+ 2.6 } $ & $ -106 _{- 11 } ^{+ 9 } $ & $ 77 _{- 55 } ^{+ 87 } $ & $ 69 _{- 47 } ^{+ 87 } $ & $ 130 _{- 68 } ^{+ 88 } $ & $ 173 _{- 50 } ^{+ 84 } $ \\
And~XXIV & $ 2.2 _{- 1.7 } ^{+ 6.0 } $ & $ 0.8 _{- 1.5 } ^{+ 7.0 } $ & $ -172 _{- 18 } ^{+ 14 } $ & $ 72 _{- 51 } ^{+ 83 } $ & $ 68 _{- 45 } ^{+ 77 } $ & $ 120 _{- 60 } ^{+ 83 } $ & $ 226 _{- 61 } ^{+ 78 } $ \\
And~XXII & $ 1.6 _{- 1.1 } ^{+ 5.4 } $ & $ 0.5 _{- 1.3 } ^{+ 6.0 } $ & $ -164 _{- 20 } ^{+ 330 } $ & $ 74 _{- 52 } ^{+ 81 } $ & $ 69 _{- 49 } ^{+ 71 } $ & $ 123 _{- 58 } ^{+ 78 } $ & $ 228 _{- 61 } ^{+ 68 } $ \\
Triangulum & $ 2.5 _{- 1.9 } ^{+ 1.4 } $ & $ -1.3 _{- 1.4 } ^{+ 1.8 } $ & $ 121 _{- 10 } ^{+ 13 } $ & $ 163 _{- 49 } ^{+ 72 } $ & $ 33 _{- 1 } ^{+ 58 } $ & $ 193 _{- 72 } ^{+ 45 } $ & $ 238 _{- 59 } ^{+ 107 } $ \\
And~VII & $ 4.0 _{- 2.3 } ^{+ 3.8 } $ & $ 3.1 _{- 2.7 } ^{+ 4.4 } $ & $ -7 _{- 5 } ^{+ 5 } $ & $ 70 _{- 49 } ^{+ 82 } $ & $ 78 _{- 58 } ^{+ 84 } $ & $ 119 _{- 65 } ^{+ 103 } $ & $ 120 _{- 65 } ^{+ 104 } $ \\
And~XXX & $ 1.3 _{- 1.0 } ^{+ 6.1 } $ & $ -0.1 _{- 1.0 } ^{+ 6.8 } $ & $ -160 _{- 18 } ^{+ 14 } $ & $ 73 _{- 52 } ^{+ 78 } $ & $ 64 _{- 43 } ^{+ 69 } $ & $ 116 _{- 57 } ^{+ 83 } $ & $ 218 _{- 61 } ^{+ 71 } $ \\
IC~10 & $ 3.6 _{- 3.2 } ^{+ 3.8 } $ & $ 2.6 _{- 3.8 } ^{+ 4.1 } $ & $ -44 _{- 8 } ^{+ 93 } $ & $ 106 _{- 83 } ^{+ 44 } $ & $ 63 _{- 39 } ^{+ 59 } $ & $ 130 _{- 65 } ^{+ 59 } $ & $ 137 _{- 49 } ^{+ 61 } $ \\
And~XXXI & $ 2.7 _{- 2.3 } ^{+ 5.2 } $ & $ 1.6 _{- 2.5 } ^{+ 5.7 } $ & $ -68 _{- 1010 } ^{+ 1009 } $ & $ 67 _{- 44 } ^{+ 86 } $ & $ 62 _{- 44 } ^{+ 77 } $ & $ 110 _{- 56 } ^{+ 94 } $ & $ 171 _{- 70 } ^{+ 102 } $ \\
And~XVI & $ 3.9 _{- 2.2 } ^{+ 4.6 } $ & $ 2.9 _{- 2.1 } ^{+ 5.4 } $ & $ 68 _{- 7 } ^{+ 8 } $ & $ 67 _{- 48 } ^{+ 79 } $ & $ 65 _{- 47 } ^{+ 70 } $ & $ 118 _{- 64 } ^{+ 79 } $ & $ 141 _{- 52 } ^{+ 74 } $ \\
And~VI & $ 4.5 _{- 4.0 } ^{+ 3.9 } $ & $ 3.4 _{- 4.1 } ^{+ 4.7 } $ & $ -40 _{- 6 } ^{+ 6 } $ & $ 70 _{- 47 } ^{+ 92 } $ & $ 69 _{- 51 } ^{+ 67 } $ & $ 115 _{- 53 } ^{+ 92 } $ & $ 123 _{- 48 } ^{+ 89 } $ \\
LGS~3 & $ 4.7 _{- 4.2 } ^{+ 4.0 } $ & $ 3.9 _{- 4.3 } ^{+ 4.4 } $ & $ -14 _{- 4 } ^{+ 4 } $ & $ 71 _{- 53 } ^{+ 75 } $ & $ 63 _{- 45 } ^{+ 66 } $ & $ 116 _{- 58 } ^{+ 79 } $ & $ 117 _{- 58 } ^{+ 79 } $ \\
And~XXVIII & $ 4.0 _{- 4.3 } ^{+ 2.9 } $ & $ 3.1 _{- 4.6 } ^{+ 3.2 } $ & $ 26 _{- 8 } ^{+ 6 } $ & $ 61 _{- 43 } ^{+ 75 } $ & $ 57 _{- 38 } ^{+ 60 } $ & $ 104 _{- 53 } ^{+ 64 } $ & $ 108 _{- 51 } ^{+ 63 } $ \\
PegasusdIrr & $ 3.1 _{- 4.9 } ^{+ 0.6 } $ & $ 2.2 _{- 5.6 } ^{+ 1.2 } $ & $ 121 _{- 10 } ^{+ 13 } $ & $ 49 _{- 41 } ^{+ 141 } $ & $ 60 _{- 49 } ^{+ 39 } $ & $ 88 _{- 50 } ^{+ 106 } $ & $ 160 _{- 46 } ^{+ 99 } $ \\
And~XVIII & $ 3.8 _{- 6.0 } ^{+ 2.7 } $ & $ 3.0 _{- 7.2 } ^{+ 2.9 } $ & $ -32 _{- 6 } ^{+ 5 } $ & $ 63 _{- 44 } ^{+ 73 } $ & $ 52 _{- 36 } ^{+ 61 } $ & $ 100 _{- 54 } ^{+ 68 } $ & $ 106 _{- 49 } ^{+ 66 } $ \\
IC~1613 & $ -1.8 _{- 0.8 } ^{+ 0.9 } $ & $ -3.4 _{- 1.6 } ^{+ 1.3 } $ & $ -68 _{- 8 } ^{+ 7 } $ & $ 60 _{- 36 } ^{+ 108 } $ & $ 61 _{- 40 } ^{+ 36 } $ & $ 88 _{- 39 } ^{+ 110 } $ & $ 117 _{- 18 } ^{+ 107 } $ \\